# Enhancing Media Literacy: The Effectiveness of (Human) Annotations and Bias Visualizations on Bias Detection

Timo Spinde [a], Fei Wu [b], Wolfgang Gaissmaier [c, d], Gianluca Demartini [e], and Helge Giese [c,f]

[a] Department of Computer Science, University of Göttingen, Göttingen, Germany.

[b] Institute for Natural Language Processing, University of Stuttgart, Stuttgart, Germany.

[c] Department of Psychology, University of Konstanz, Konstanz, Germany.

[d] Cluster for the Advanced Study of Collective Behavior, University of Konstanz, Konstanz, Germany.

[e] School of Electrical Engineering and Computer Science, The University of Queensland, Brisbane, Australia.

[f] Heisenberg Chair for Medical Risk Literacy and Evidence-Based Decisions, Charité – Universitätsmedizin Berlin, Berlin, Germany

**Corresponding Author**

Correspondence concerning this article should be addressed to Timo Spinde, University of Göttingen, Papendiek 14, 37073 Göttingen, Germany; e-mail: t.spinde@media-bias-research.org

# Enhancing Media Literacy: The Effectiveness of (Human) Annotations and Bias Visualizations on Bias Detection

## ABSTRACT

Marking biased texts is a practical approach to increase media bias awareness among news consumers. However, little is known about the generalizability of such awareness to new topics or unmarked news articles, and the role of machine-generated bias labels in enhancing awareness remains unclear. This study tests how news consumers may be trained and pre-bunked to detect media bias with bias labels obtained from different sources (Human or AI) and in various manifestations. We conducted two experiments with 470 and 846 participants, exposing them to various bias-labeling conditions. We subsequently tested how much bias they could identify in unlabeled news materials on new topics. The results show that both Human ($t(467) = 4.55$, $p < .001$, $d = 0.42$) and AI labels ($t(467) = 2.49$, $p = .039$, $d = 0.23$) increased correct detection compared to the control group. Human labels demonstrate larger effect sizes and higher statistical significance. The control group ($t(467) = 4.51$, $p < .001$, $d = 0.21$) also improves performance through mere exposure to study materials. We also find that participants trained with marked biased phrases detected bias most reliably ($F(834,1) = 44.00$, $p < .001$, $\eta^2_{part} = 0.048$). Our experimental framework provides theoretical implications for systematically assessing the generalizability of learning effects in identifying media bias. These findings also provide practical implications for developing news-reading platforms that offer bias indicators and designing media literacy curricula to enhance media bias awareness.

## Introduction

News media play a central role in informing the public about crucial societal issues. In areas where audiences do not have direct knowledge or experience, the audiences become particularly reliant on the media that transmit the information (Happer & Philo, 2013).News media often introduce biases by simplifying complex scientific or social issues to create a storyline that engages the audience (Henderson & Green, 2020). The term "media bias" refers to such slanted news coverage or other biased media content (Hamborg et al., 2019; Spinde, 2021) that favors a particular perspective, ideology, or result (Williams, 1975). During the news production process, bias can be introduced in various ways, such as through journalists' choices of biased words and framing of a topic, or editorial decisions (Spinde et al., 2024).

Media bias significantly impacts the public's knowledge of specific issues and decision-making processes, especially when news consumers are unaware of the extent of the bias (Druckman & Parkin, 2005; Eberl et al., 2017; Spinde et al., 2024). For instance, during the pandemic of COVID-19, Dhanani & Franz (2020) found those who had greater knowledge of the pandemic were more likely to be critical of biased media content when formulating their own opinions, while those who relied only on certain news sources were more likely to be influenced unconsciously. Furthermore, biased media content can lead to polarization in society and foster the audience's distrustful attitude toward the authorities (Dhanani & Franz, 2020). Media bias also impacts other domains such as politics (Bernhardt et al., 2008; Eberl et al., 2017; Islam, 2008), economics (McCarthy & Dolfsma, 2014), environment (Henderson & Green, 2020), and social safety issues (Sugimoto et al., 2013). These observations highlight a need for strategies that mitigate the negative effects of media bias, especially in cases of controversial topics that are societally relevant.

Furthermore, the rapid news production cycle on digital platforms has amplified the amount of biased content exposed to news consumers (Iandoli et al., 2021; Terren & Borge-Bravo, 2021).

Customized news recommendation systems also contribute to echo chambers (Auxier & Vitak, 2019), where news consumers are more appealed to read news aligning with their opinions, and their biased perspectives are further reinforced (Spinde et al., 2024). As the volume of digital information continues to grow, obtaining an overview of bias across media outlets is only feasible through automated solutions (Wessel et al., 2023).

Previous research has explored mitigating media bias through visualizations that highlight biased text (Spinde et al., 2020, 2022) or present diverse viewpoints to encourage a more balanced news consumption (Joris et al., 2024; Mattis et al., 2024; S. Munson et al., 2021; Paramita et al., 2024; Park et al., 2009, 2012; Rieger et al., 2024). While these studies show that visualizations can help consumers recognize bias, they lack a systematic approach to assess the learning effects after removing visual aids and test bias detection on unfamiliar topics. Our research addresses the gap by dividing the experiment into training and testing phases. In the training phase, participants view articles with multiple linguistic bias visualizations. In the testing phase, they identify biased content in plain articles on different topics, allowing us to examine whether the learning effect transfers to new topics and persists post-visualization removal.

Research also explored using natural language processing (NLP) and machine learning to automatically detect media bias (He et al., 2021; Hube & Fetahu, 2019; Lee et al., 2022; Lei et al., 2022; Liu et al., 2021; Pryzant et al., 2020; Spinde, Plank, et al., 2021). While their results indicate that scalable, automated methods are possible for handling the amount of daily news content (Hamborg et al., 2019; Spinde et al., 2024), the effectiveness of automatically generated visualizations in enhancing humans' abilities to identify media bias remains untested. Prior visualization experiments rely on manual annotations, limiting scalability. Our study aims to bridge this gap by evaluating the potential of automated labels for media bias visualization and training. We formulate our research questions (RQ) as follows:

- RQ1: How effective are machine-generated bias labels in training human participants to detect media bias, compared to Human labels and a control group without a media bias training?

- RQ2: How generalizable are the learning effects of media bias to new topics and materials post-visualization removal?

- RQ3: Which visualization strategies are most effective in training participants to detect media bias?

- RQ4: Do participants' backgrounds, such as political inclinations, interact with different visualizations to jointly influence their learning effects?

We address RQ1 and RQ2 in Study 1, with a focus on the training efficacy of human-annotated versus machine-generated labels and the learning effects post-exposure on detecting bias in single sentences. We address RQ2, RQ3, and RQ4 in Study 2, where we test the efficacies of new visualization strategies at both sentence and article levels. We preregistered our hypotheses at https://osf.io/q8rcu/ and https://osf.io/95ht6/, where we also make all data and code publicly available. Through these studies, we make the following contributions:

1. We demonstrate the potential of automated bias labels to enhance participants' ability to detect bias. While participants trained with Human labels outperform those trained with machine labels, automated labels still significantly improve media bias awareness, highlighting their potential as an alternative to Human labels in bias visualizations.

2. We propose a systematic approach to measuring the generalizability of media bias learning effects. Our findings show that participants continue to detect bias in the test phase, suggesting the learning effects persist and can generalize to plain news content of unfamiliar topics.

3. We assess the efficacy of new visualization strategies and find that bias highlights at the phrase level are the most effective in raising media bias awareness.

4. We examine the interaction between visualization strategies and participants' political orientation, showing that the learning effects weaken when both phrasal bias and politicized word use are highlighted. This supports previous findings that echo chambers remain challenging to overcome, even with visual aids.

## Related Work

Media bias can materialize in various ways, such as linguistic choices, editorial decisions, and even the broader socio-political environment in which media operates (Spinde et al., 2024). This research focuses on linguistic bias, which is the basis for the phrase- and sentence-level visualized labels we tested. Linguistic bias refers to word choices that reinforce stereotypes or specific perceptions of events, groups, or individuals (Beukeboom & Burgers, 2017) and is evident in individual words, phrases, and sentence structure (Hinterreiter, Spinde, et al., 2024).

Research on mitigating media bias has explored various strategies for raising news consumers' awareness of media bias. One main strategy focuses on motivating readers to engage with diverse perspectives, while others focus on developing bias-indicating visualizations at the word and sentence levels. The following sections will review these strategies, examining their methods, effectiveness, and their shortcomings.

### Aggregation of Diverse News Perspectives

One active line of research focuses on nudging readers toward more balanced news consumption by aggregating diverse perspectives or using ranking algorithms to increase news diversity in recommender systems and search engine results pages (Joris et al., 2024; Mattis et al., 2024; S. Munson et al., 2021; Paramita et al., 2024; Park et al., 2009, 2012; Rieger et al., 2024). However, these studies did not measure users' awareness of media bias before and after using the proposed methods, leaving the effectiveness of these methods in raising individuals' media bias awareness unknown. In contrast, Bhuiyan et al. (2023) explored how annotating the similarity and dissimilarity of articles affects annotators' perceptions of quality and credibility. Spinde et al. (2020) found that visualizing diverse perspectives did not significantly affect bias perception. However, both studies assessed participants' perceptions of bias through questionnaires, and their ability to detect bias in plain news articles remains unclear.

**Bias-Awareness Visualizations**

Spinde et al. (2020) investigated how highlighting different content types (key facts, framing effects, biased language) affects human bias detection in news articles, finding that highlighting biased language was the most effective. In a follow-up study (Spinde et al., 2022), they evaluated additional bias indicators, discovering that forewarnings and biased language annotations with explanations were effective, with annotations being the most impactful, while political classification had no effect. However, these studies have measured bias awareness through attitude shifts (Spinde et al., 2020) or querying their perception of bias of the presented materials (Spinde et al., 2022), but these findings are limited in scalability due to reliance on manual annotations and subjective questionnaire items.

Building on the finding that highlighting biased language can raise bias awareness (Spinde et al., 2020, 2022), Hinterreiter, Wessel, et al. (2025) introduced *NewsUnfold*, a media bias dataset annotated by crowdsourcers who provided feedback on AI-generated bias-indicating visualizations, significantly enhancing the baseline dataset quality with a 26.31% improvement in inter-annotator agreement. Another media bias dataset, *News Ninjia* (Hinterreiter, Spinde, et al., 2024), achieves comparable quality to expert annotations by providing bias-indicating visualization feedback to crowdsourcers. However, these studies did not evaluate participants' ability to detect media bias, as this was not the focus of the research, and the short- and long-term learning effects remained unclear (Hinterreiter, Spinde, et al., 2024; Hinterreiter, Wessel, et al., 2025).

In summary, there is a lack of a systematic approach to evaluating the generalizability of learning effects across new topics or their persistence after removing visualization aids. Our research addresses these gaps by dividing the experiment into training and testing phases to assess learning effects after removing the visualizations, while also evaluating the effectiveness of automatically generated bias labels.

## Automatic Media Bias Detection

Given the vast amount of online news content, relying solely on manual annotations is impractical. To address this, researchers have suggested automated labeling as a scalable alternative (Wessel et al., 2023). Natural language processing (NLP) techniques can be used to automate bias detection, enabling large-scale systems for bias analysis and mitigation (Hinterreiter, Wessel, et al., 2025; Spinde et al., 2024; Wessel et al., 2023). Although these models have shown promising results, their training relies on manual annotations by experts or crowdsourcers. For instance, Spinde, Plank, et al. (2021b) achieved classifier performance with a macro F1-score of 0.804 using the Bias Annotation By Experts (BABE) dataset. Despite the dataset's high quality and strong inter-rater agreement, its limited coverage of topics and timeframes restricts its scalability.

However, it remains unclear whether automated labels are as effective as human annotations in training classifiers to detect media bias and in helping readers become more aware of biased content. Our study aims to bridge this gap by comparing the effectiveness of automated labels versus human-made labels in improving media bias awareness among readers.

## Study 1

### Participants

We recruited 512 participants on Prolific taking part in a 12-minute survey with a reimbursement of 1.50£. We used Prolific's feature of assigning a representative sample of sex, age, and ethnicity. The platform uses census data from the US Census Bureau or the UK Office of National Statistics to divide the sample into subgroups with the same proportions as the national population. The survey was published on 11th November 2021. Based on these preregistered exclusion criteria, we used data from 470 for our analyses (47.9% men, 49.5% women, 2.4% other; $M_{age}$ = 31.2, $SD_{age}$ = 11.4, $Md_{education}$ = Bachelor). On average, participants indicated a political orientation score of 3.4 ($SD$ = 2.4; 0 = very liberal, 10 = very conservative). The analyzed sample size was sufficiently powered with a power of .95 to find small to medium effects of $f$ = .18.

**Bias labels**

We develop teaching visualizations for biased sentences in the training phase using both Human-annotated and AI-generated labels that indicate whether a sentence contains biased language. For Human labels, we directly use the binary sentential bias labels (Biased vs. Non-biased) from the Bias Annotations By Experts (BABE) dataset (Spinde, Plank, et al., 2021). The BABE corpus contains 3,700 sentences from diverse U.S. news outlets, balanced across topics, with bias labels annotated by five trained media experts at both the word and sentence levels. Using Krippendorff's metric α, BABE demonstrates higher inter-annotator agreement (α = 0.40) and better annotation quality compared to previous datasets (Spinde, Rudnitckaia, et al., 2021). We extract 46 sentences from BABE, evenly split between training and testing phases, to obtain a representative collection regarding the underlying political distribution of news outlets, as shown in Fig. 1.

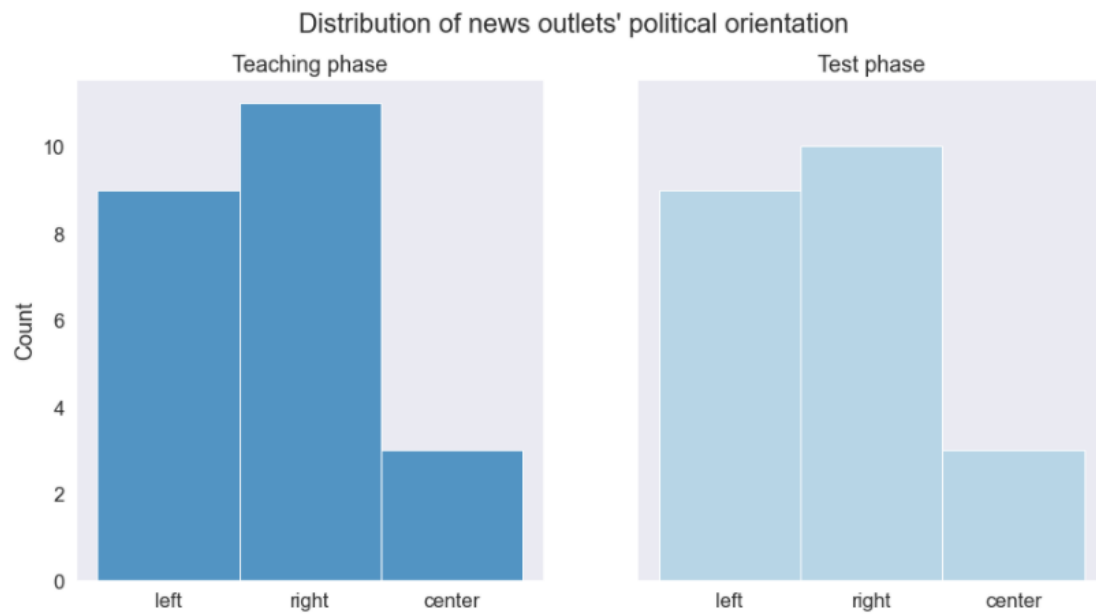

For AI labels, we first build on recent research in automated bias detection (Pryzant et al., 2020; Spinde, Plank, et al., 2021) by training the state-of-the-art neural-based language model RoBERTa with a supervised approach on a large news corpus of biased and neutral text. We then use this model, BiasRoBERTa, which achieves an F1 score of 0.814 on the BABE corpus, to classify the 23 sentences as Biased or Non-biased for our AI-generated labels. Based on the Human labels from BABE and the AI-generated labels, we create simple visualizations highlighting sentence-level bias.

**Method**

After informed consent, each participant was randomly assigned a condition and introduced to the concept of bias, and the potential training information according to condition. Then, all participants were presented with the same 23 training sentences with markings according to the condition. Subsequently, all participants were presented with 23 test sentences without any markings, regardless of condition. Finally, the participants provided general demographic information and indicated that their data could be trusted before being debriefed and compensated.

**Experiment design**

The experiment followed a 3-*label* (Human label/AI label/control) between × 2 *phases* (training/test) within design. In both the Human label and the AI label condition, training phase sentences were highlighted in yellow if the respective source (Human or AI) in the BABE dataset (Spinde, Plank, et al., 2021) classified it as biased, while in the control condition and the test phase, all sentences were presented without any markings.

For each sentence, we asked participants: "Please tell us what you think about the sentence…" to assess their perception of bias. Participants then responded on a 6-point Likert scale, ranging from "strongly disagree" to "strongly agree," to the statement: "In my opinion, this sentence is biased," as shown in Fig. 2. An attention check was included in both the training and testing phases, instructing participants to select a specific option in the bias rating question.

**Sentence:**
*"Ocasio-Cortez used this weekend's news cycle to continue highlighting the evils of wealth inequality, and to draw attention to serious policy fixes for the problem."*

Please tell us what you think about the sentence...

| | Strongly disagree | Disagree | Somewhat disagree | Somewhat agree | Agree | Strongly agree |
|---|---|---|---|---|---|---|
| In my opinion, this sentence is biased. | ○ | ○ | ○ | ○ | ○ | ○ |

*Fig. 2. A sample question from the survey presenting a sentence and asking for the subject's media bias perception.*

## Data analysis

The bias perception rating of each sentence was binarized to compare it with the bias classifiers and obtain an accuracy score. Accordingly, the F1 score as a measure for classifying accuracy was computed for the training and test phase of each participant taking the human classification of the sentences according to the BABE (Spinde, Plank, et al., 2021) dat dataset as the comparator. To address RQ1 and RQ2, we propose two null hypotheses for testing:

The bias perception rating of each rating was binarized to compare it with the bias classifiers and obtain an accuracy score. Accordingly, the F1 score as a measure for classifying accuracy was computed for training and test phase of each participant taking the human classification of the sentences according to the BABE (Spinde, Plank, et al., 2021) dataset as comparator. To address RQ1 and RQ2, we propose two null hypotheses for testing:

**Null Hypothesis 1:** AI-generated labels and Human labels do not enhance participants' ability to detect media bias compared to a control group that receives no training.

**Null Hypothesis 2:** After the training sessions with different visualizations end, participants' ability to detect media bias does not generalize to new topics and materials during the testing session.

We started with analyzing descriptive statistics on the participants' labeling accuracy over all groups and survey phases. We used a 3-*label* (Human/AI/control) × 2-*phase* (training/test) mixed ANOVA to test how achieved labeling accuracy as F1 scores were affected by the different types of training using the R package afex. The ANOVA was followed by (Sidak corrected) post-hoc test using the package emmeans.

**Results & Discussion**

Overall, the discovered effect of *label* ($F(467,2) = 10.36$, $p < .001$, $\eta^2_{part} = 0.042$) denotes that both the Human label ($t(467) = 4.55$, $p < .001$, $d = 0.42$) as well as the AI label ($t(467) = 2.49$, $p = .039$, $d = 0.23$) improved correct detection of media bias compared to the control. Therefore, Null Hypothesis 1 should be rejected. Likewise, classifications improved from the training to testing phase ($F(467,1) = 32.38$, $p < .001$, $\eta^2_{part} = 0.065$; Fig. 3), suggesting that Null Hypothesis 2 should also be rejected.

As predicted, these effects were qualified by a *label × phase* interaction ($F(467,2) = 5.06$, $p = .007$, $\eta^2_{part} = 0.021$): Unsurprisingly, participants with Human label training had better accuracy than the ones with AI label ($t(467) = 3.17$, $p = .0048$, $d = 0.29$) and controls ($t(467) = 5.32$, $p < .001$, $d = 0.49$) in the training phase ($F_{label\backslash train}(467,2) = 14.34$, $p < .001$, $\eta^2_{part} = 0.06$), because their accuracy was directly evaluated by the Human labels. While both control ($t(467) = 4.51$, $p < .001$, $d = 0.21$) and AI label ($t(467) = 4.67$, $p < .001$, $d = 0.22$) improved accuracy in the test phase and the Human label accuracy did not change systematically ($t(467) = 0.69$, $p = .4927$, $d = 0.03$), Human labels remained superior to the control in the test phase ($t(467) = 2.68$, $p = .0228$, $d = 0.25$; $F_{label\backslash test}(467,2) = 4.12$, $p = .0168$, $\eta^2_{part} = 0.02$). AI labels and the control could only be marginally differentiated in the test phase ($t(467) = 2.23$, $p = .0768$, $d = 0.21$). The effectiveness of the training was independent of the political identification of the trained participant (all political identification effects: $F \leq 2.66$, all $p \geq .104$, all $\eta^2_{part} \geq 0.006$).

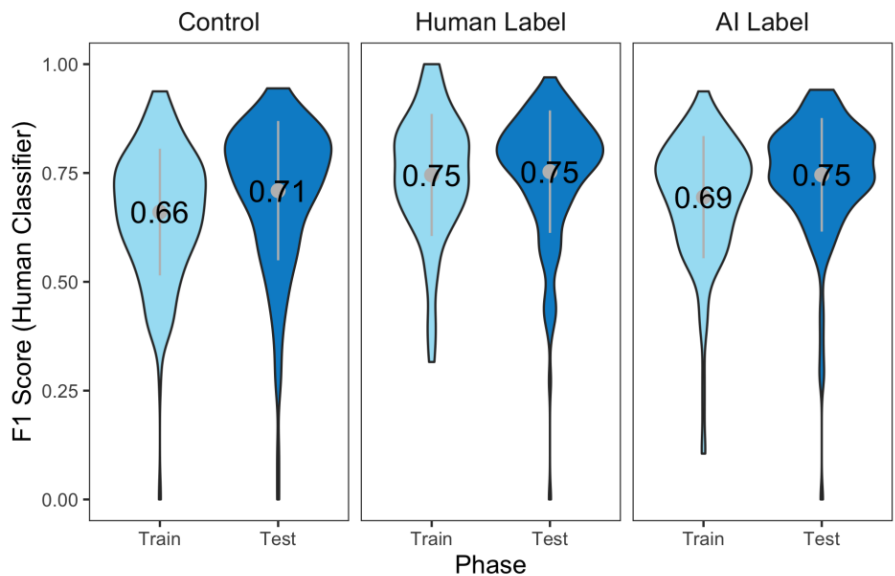

*Fig. 3.* *Distribution of participants' F1 scores over all survey groups and phases.*

## Study 2

Generally, training that exposes participants to media-biased phrases or sentences proved effective to improve the ability to detect biases in subsequent texts–even if annotations themselves may not be fully accurate. These effects go beyond mere exposure and attention to the problem by posing questions (Pennycook et al., 2021), though learning effects across time also indicate that these effects of attention do exist. Furthermore, by marking specific phrases as biased without underlining the potential political motivation for it, training was effective across political isles. Still, while Study 1 is a general proof of concept that media bias training can generally be successful in small to medium effect sizes, it was confined to single sentences and rather simple feedback training. In Study 2, this training approach is generalized to detecting biased language in an article. Specifically, what remains unclear in this broader context is how biased language annotations may be optimized to train bias detection: Are trainings still possible when confined to single, longer texts, or are the effects merely found on the level of single sentences? Is it sufficient to just generally flag a sentence in an article that is supposed to contain a bias—as in Study 1, or do news consumers additionally profit from specifically highlighting the biased words and suggestions of how the sentences may be phrased more neutrally? Furthermore, does a political contextualization of an article help to learn to detect biased language by illustrating which phrases are used in a specific political milieu (D'Alonzo & Tegmark, 2022).

**Participants**

The study was sampled on the sampling platform Prolific. Qualifications entailed equal gender distribution, USA residency, fluency in English, a minimum of 50 and a maximum of 10000 completed tasks, and an approval rate of more than 95% via Prolific settings. Sampling occurred on January 15, 2022, and workers were compensated with 2.25£. A total of 1121 participants were recruited, of which $N_{full}$ = 846 participants (434 female, 391 male, 16 other, 5 not stated; age: $M$ = 35.15, $SD$ = 13.28) provided data used in the analyses. This sample is 95% powered to detect small effects of at least $f$ = .13 for the preregistered comparisons.

**Method**

After informed consent and assessment of demographic information, participants had to indicate their political orientation. Then they were assigned a feedback condition and were asked to complete the study in two phases. In the first phase, the training phase, they were presented with a tutorial introducing them to the feedback visualizations according to condition, followed by a presentation of two quasi-randomly selected articles, for which they were asked to determine which of the two was less biased. In a second phase, the test phase, all participant received the same materials: a tutorial explaining how to mark biases by clicking on the sentences and phrases they deemed biased, followed by a blank test article, for which they were asked to indicate all bias. Finally, participants were debriefed and compensated.

**Materials and design**

The manipulated feedback in the training phase followed a 3-*biased sentence labeling* (absent/simple/advanced) × 2 *biased phrase labeling* (absent/present) × 2 *politicized phrase labeling* (absent/present) between design. Fig. 4 shows the task screen entailing the all-label conditions. For each article, politicized phrases were obtained from D'Alonzo & Tegmark (2022) topic-related discriminative phrase categorization and min-max normalized per topic, biased phrases and sentences were determined by two raters agreeing on the classification following the annotation guidelines by Spinde, Plank, et al. (2021b). In the label conditions, biased phrases were underlined, politicized phrases were highlighted with a dotted underline, and the combination of both with a dashed underline. Biased sentences were highlighted in the respective conditions with a grey background color. Additional information (i.e., the left-right bias of politicized phrases and/or a synonymous unbiased sentence) was displayed above the article in the so-called "Analysis Bar" for conditions entailing these advances communications when selected by a mouse click (politicized phrases condition, advanced biased sentences condition).

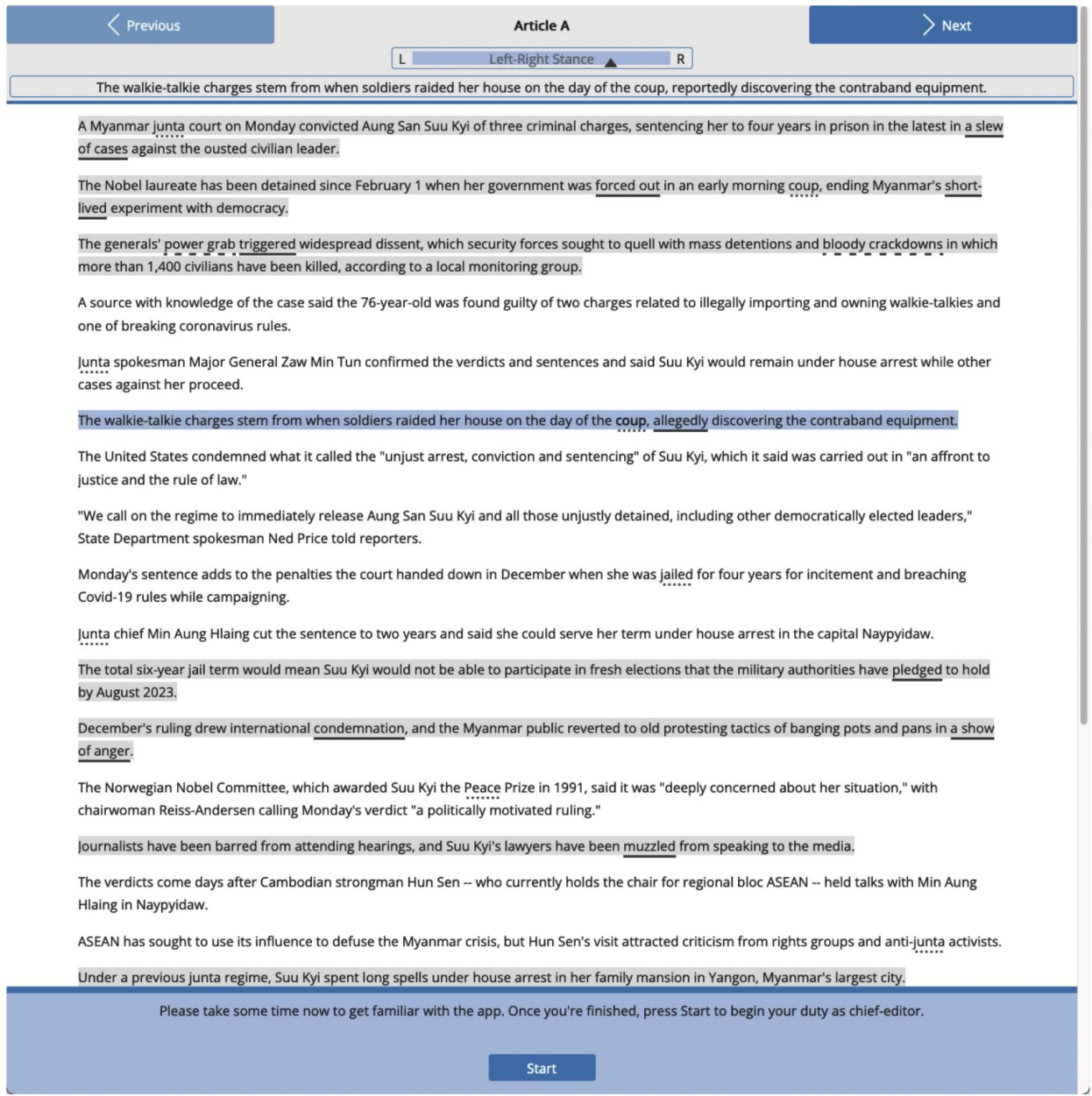

*Fig. 4.  Screenshot of the online experiment.*

For both phases, articles about the same topic and political ratings were collected using GroundNews. Following findings that the congruence of the political attitude of an article and a participant may affect its perception (Spinde et al., 2022), the articles for the training phase were selected across political isles. Conversely, the test article was chosen to be neutral and uncontroversial to provide similar preconditions for all participants. Accordingly, the training tutorial article included two biased articles about Myanmar's former president Aung San Suu Kyi. The training task consisted of two biased articles about weapon restrictions in the USA, specifically the trial of Kyle Rittenhouse, sampled from two left, center, and right media sources. To grant comparable stimulus training conditions, all articles were algorithmically modified based on the BABE dataset (Spinde, Plank, et al., 2021) to contain 3.33%, 6.66%, or 10% of biased words, respectively. In addition, longer articles were manually trimmed with the lowest loss of information possible to contain roughly the same number of words. A clustered random assignment to the articles ensured that one article was presented in a less biased version than the other. The test phase tutorial used an article about the shutdown of an independent Hong Kong news portal. In the test trial, all participants annotated an article about the launch of the James Webb telescope. For detailed information about the articles, see the materials at osf.io.

In the training phase, participants had to determine which of the two presented articles they deemed more biased. In the test phase, participants were asked to indicate biased phrases in the test article by highlighting them with a double click.

**Data analysis**

The $F1$ score was computed for the test article on sentence level. The classification of biased sentences was obtained from Spinde, Plank, et al. (2021b). Accordingly, marked biased sentences were counted as hits, marked unbiased sentences as false alarms, unmarked biased sentences as misses, and unmarked unbiased sentences as correct rejections. Please note that $F1$ scores are lower than in Study 1 because participants were not forced to react to all sentences yielding a higher rate of misses.

To address RQ2, RQ3, and RQ4, we propose the following three null hypotheses for testing:

| |
|---|
| **Null Hypothesis 2:** After the training sessions with different visualizations end, participants' ability to detect media bias does not generalize to new topics and materials during the testing session. **Null Hypothesis 3:** All visualization strategies show no difference in their effectiveness in training participants to detect media bias. |

> **Null Hypothesis 4:** Participants' backgrounds, such as political inclinations, interact with different visualizations and jointly influence their learning outcomes.

A 3 *biased sentence labeling* (absent/simple/advanced) × 2-*biased phrase labeling* (absent/present) × 2 *politicized phrase labeling* (absent/present) between-subject ANOVA on the extracted F1 scores was performed to test whether the feedback affected participant's ability to identify biased sentences. Effects of *biased sentence labels* were followed by simple contrast (Bonferroni-corrected) and interactions by simple effects analyses using the R package emmeans.

To test how the feedback affected the ability to correctly identify the more biased article of the two presented in the training phase, a 3-*biased sentence labeling* (absent/simple/advanced) × 2-*biased phrase labeling* (absent/present) × 2-*politicized phrase labeling* (absent/present) logistic regression with the article choice (higher 0/lower biased article 1) was performed as the criterion and effect-coded factors.

To test how political identification affected the feedback training, this design was extended to a GLM adding political orientation (scaled from −50 (left) to +50 (right)), article orientations, and all interactions. Article orientations were formed as an average of the orientation of both presented articles in the training phase coded -1 for left, 0 for neutral, and 1 for right articles. All factors were effect coded and all interactions were included.

The analyses were preregistered under OSF.IO/3VDCH. Due to a technical redirection error after experiment completion, data of about three times the preregistered sample size was accidentally acquired. Analyses reported use all the data using the preregistered screening criteria and additionally excluding one individual taking implausibly long (more than 9 SDs longer than the mean), as the data of the targeted sample are indiscernible from the oversampled one. With this effective sample size of 846, we are powered with a .95 probability to find significant small effects of at least $f = .13$.

**Results & Discussion**

***Biased sentence identification in the test task***

Overall, participants were better able to correctly identify biased sentences in the test phase, when they were trained with a *biased sentence* ($F(834,2) = 5.58$, $p = .004$, $\eta^2_{part} = 0.012$) and *biased phrase* labels ($F(834,1) = 44.00$, $p < .001$, $\eta^2_{part} = 0.048$). This suggests Null Hypothesis 2 should be rejected, as in Study 1.

However, F1-scores were decreased by *politicized phrase* labels ($F(834,1) = 4.16$, $p = .042$, $\eta^2_{part} = 0.005$; see Fig. 5). While a negative effect of politicized phrases in this context is somewhat surprising and novel, it is in line with earlier findings indicating that the political dimension is somewhat unrelated and may even obscure the ability to detect biases (D'Alonzo & Tegmark, 2022).

The interaction of both successful trainings (*biased sentence × biased phrase*: $F(834,2) = 4.16$, $p = .002$, $\eta^2_{part} = 0.01$) further illustrated that *biased phrase* labelling was sufficient to evoke effects: Simple and advanced biased sentence labels in the training phase were only superior in F1 accuracy compared to control, when no additional biased phrase labels were presented ($t_{simple\ vs.\ absent\backslash no\ biased\ phrases}(834) = 4.28$, $p < .001$, $d = 0.51$; $t_{advanced\ vs.\ absent\backslash no\ biased\ phrases}(834) = 4.28$, $p < .001$, $d = 0.47$; $t_{simple\ vs.\ absent\backslash biased\ phrases}(834) = 0.22$, $p = 1$, $d = 0.03$; $t_{advanced\ vs.\ absent\backslash biased\ phrases}(834) = 0.50$, $p = 1$, $d = 0.06$).

On the other hand, *biased phrase* labeling always increased performance regardless of *biased sentence* labeling (all $t(834) \geq 2.22$, all $p \geq .027$, all $d \geq 0.26$). In addition, there seemed to be a marginal trend that *biased phrase* labeling was to some degree less effective when the training also contained *politicized phrase* labels ($F(834,1) = 3.42$, $p = .065$, $\eta^2_{part} = 0.004$). No other effect on sentence-level F1-scores emerged (all $p \geq .065$, all $\eta^2_{part} \leq 0.004$). Overall, the results suggest that Null Hypothesis 3 should be rejected, as different visualization strategies demonstrated varying effectiveness in training participants to detect media bias.

Testing for potential mitigation by political views of both the training material and the trainees, we found that effects of *biased phrase* labels in the training phase hold controlling for political ideologies ($F(798,1) = 29.100$, $p < .001$, $\eta^2_{part} = 0.033$). Still, more conservative participants were generally less accurate ($F(798,1) = 6.48$, $p = .011$, $\eta^2_{part} = 0.007$) and also decreased the effects of *biased sentence* labeling (*biased sentence* × *political affiliation*: $F(798,2) = 5.43$, $p = .005$, $\eta^2_{part} = 0.01$), so that the effects were not significantly present in politically neutral participants *(biased sentence $F(798,2) = 1.04$, $p = .353$, $\eta^2_{part} = 0.002$)*. This indicates that participants' political inclinations did interact with the visualizations and affect learning outcomes, particularly for conservative participants, suggesting that Null Hypothesis 4 should be accepted. There were no further effects of political viewpoints (all $p \geq .168$, all $\eta^2_{part} < 0.01$).

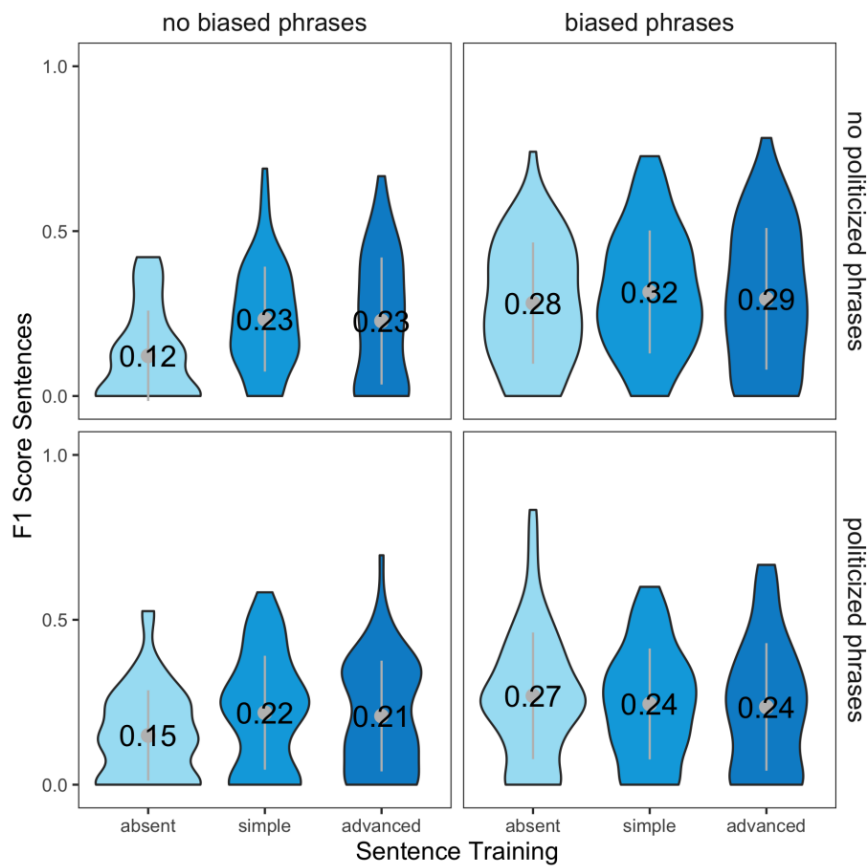

**Fig. 5.** *Distribution of participants' F1 scores over all survey groups during the testing phase.*

### Biased article selection in training

Participants only correctly identified the higher biased article of the training set more consistently (62.7% vs. 70.8%) when being provided with *politicized phrase* labeling ($\chi^2(1) = 6.19$, $p = .0129$, $OR = 1.46$) with no other effect being significant (all $p \geq .086$). This illustrates that participants were directly affected in their ranking decision by the political feedback of the experimenter on the degree of bias.

### General Discussion

Training sessions that expose participants to media-biased phrases or sentences have consistently demonstrated efficacy in enhancing the ability to detect biases in subsequent texts—even when the annotations themselves might not be entirely accurate. Notably, marking specific phrases as biased, without emphasizing potential political motivations, proves effective across the political spectrum. The results indicate that, although media bias is a complex concept, visualizations highlighting phrasal- and sentential-level biased content act as a nudge, encouraging readers to think more critically about bias in the news. While pinpointing every biased instance within a comprehensive text can be challenging—possibly due to diminished motivation to identify all instances—accuracy at the sentence level shows an increase by up to an F1-Score of .08, signifying a medium-sized effect. Crucially, our study shows that while automated labels are not as accurate or effective as Human labels, they can still contribute to raising news consumers' awareness of media bias to some extent. Automated labels can be feasibly scaled to handle the vast amount of online content, making them especially useful in cases where human annotations are unavailable. It's imperative to clarify that the observed training effects aren't merely a byproduct of heightened attention to bias caused by the tutorial on bias annotation. Control groups in Study 1 are also provided with identical instructions to rate sentences and articles, yet the training effect observed in this group suggests that minimal exposure effects coexist alongside the more pronounced training effects of feedback labels.

While we found that at the sentence level, learning effects can be generalized to new, unannotated news materials with unseen topics to detect biased language, the annotations do not consistently result in the accurate classification of entire articles. This discrepancy might stem from a less defined notion of bias at the article level, allowing political inclinations to exert a more pronounced influence when comparing articles. This theory aligns with the lower performance of politicized language highlighting, consistent with prior findings on the impact of visualizing political affiliations (Spinde et al., 2022).

Given that the articles used in the study are restricted to specific topics or isolated sentences, the generalizability of the learning effects to a broader set of articles might be constrained. Addressing this limitation is challenging due to the practical constraints of sample size—a challenge shared by many studies in the media bias domain. Another area of concern is the longevity of the training effects, which the current study design did not cover.

Since the highlighting of biased language depends on context and can only be scaled through AI-driven detection, further research should focus on improving automated highlighting to ensure its accuracy in identifying bias in new news content. Even more, after demonstrating the effectiveness of our approach, the next question is how to build applications that provide these benefits for news consumers.

Our findings hold great practical implications not only for developing news-reading platforms that offer bias indicators for everyday news consumption, but also for use in schools and media literacy education. As citizens with diverse perspectives on societal issues help strengthen democracy, this research can serve as a first step toward developing strategies, tools, and resources for promoting media literacy and bias-aware news consumption. However, further research is needed to explore how these insights can be effectively applied to real-world news consumers and integrated into educational content.

## Conclusion

This study addresses the gap in systematically assessing the generalizability of media bias training by conducting two experiments to evaluate participants' ability to identify bias in plain news articles on new topics after training with various bias labels. Our findings demonstrate that both Human and AI-generated bias labels can effectively enhance media bias awareness, and that phrasal labeling proves to be the most effective visualization, regardless of political tendencies. Notably, highlighting politicized language and inclinations may have negative effects, aligning with previous findings.

Compared to prior research, this study uniquely examines the differential effectiveness of Human and AI-generated labels and various bias-indicators across varied training conditions, offering insights into optimal labeling strategies. Theoretical implications include proposing empirical frameworks for evaluating the generalizability of bias-awareness training to new contexts, providing a systematic approach to measure learning effects across diverse content.

Practically, the findings suggest the potential of integrating bias indicators into news-reading platforms and media literacy curricula to enhance consumers' ability to detect bias. A key gap to address next is translating these findings into real-world applications, such as developing tools for bias-aware news consumption or incorporating bias-awareness training into educational programs. Future research should also refine the accuracy of AI-generated labels when applied to new content, ensuring they adapt to the dynamic nature of news and consistently raise readers' awareness of bias.

**CRediT authorship contribution statement**

Timo Spinde: Conceptualization, Methodology, Data Curation, Writing - Original Draft, Writing - Review & Editing. Fei Wu: Writing - Original Draft, Writing - Review & Editing; Wolfgang: Funding Acquisition; Gianluca: Writing - Review & Editing, Funding Acquisition; Helge Giese: Conceptualization, Methodology, Writing - Original Draft, Funding Acquisition

**Funding**

This work was supported by the Hanns-Seidel Foundation, the German Academic Exchange Service (DAAD), and the University of Queensland Visiting Student Scheme. None of the funders played any role in the study design or publication-related decisions.

## Acknowledgments

We would like to thank Smi Hinterreiter, Anna Bahß, David Krieger, and Tilman Hornung for their efforts in organizing data assessment for the studies and Bela Gipp for his scholarly advice.

### Declaration of generative AI and AI-assisted technologies in the writing process

During the preparation of this work the authors used ChatGPT in order to proofread the document. After using this service, the authors reviewed and edited the content as needed and take full responsibility for the content of the publication.